\documentclass[%
 reprint,
superscriptaddress,
 amsmath,amssymb,
 aps,
]{revtex4-2}

\usepackage{siunitx}
\usepackage{graphicx}
\usepackage{dcolumn}
\usepackage{bm}
\usepackage{hyperref}
\usepackage{cleveref}
\usepackage{xcolor}
\usepackage{booktabs}
\usepackage{tikz}
\usepackage{tikz-3dplot}
\usetikzlibrary{arrows.meta,matrix}
\usepackage{pgfplots}
\tdplotsetmaincoords{60}{125}
\pgfplotsset{compat=1.16}
\usepackage{overpic}

\usepackage{ulem}

\begin{document}

\preprint{APS/123-QED}

\title{Miscible fluids patterning and micro-manipulation using vortex-based single-beam acoustic tweezers}

\author{Samir Almohamad}
\thanks{These authors contributed equally.}
\affiliation{%
Univ. Lille, CNRS, Centrale Lille, Univ. Polytechnique Hauts-de-France, UMR 8520 -
IEMN - Institut d’Electronique de Microélectronique et de Nanotechnologie, F-59000
Lille, France
}%

\author{Gustav K. Modler}
\thanks{These authors contributed equally.}
\affiliation{%
 Department of Physics, Technical University of Denmark, DTU Physics Building 309,\\ 2800 Kongens Lyngby, Denmark
}%

\author{Ravinder Chutani}
\affiliation{%
Univ. Lille, CNRS, Centrale Lille, Univ. Polytechnique Hauts-de-France, UMR 8520 -
IEMN - Institut d’Electronique de Microélectronique et de Nanotechnologie, F-59000
Lille, France
}%

\author{Udita U. Ghosh}
\affiliation{%
Univ. Lille, CNRS, Centrale Lille, Univ. Polytechnique Hauts-de-France, UMR 8520 -
IEMN - Institut d’Electronique de Microélectronique et de Nanotechnologie, F-59000
Lille, France
}%

\author{Sarah Cleve}
\email{sarah.cleve@univ-lille.fr}
\affiliation{%
Univ. Lille, CNRS, Centrale Lille, Univ. Polytechnique Hauts-de-France, UMR 8520 -
IEMN - Institut d’Electronique de Microélectronique et de Nanotechnologie, F-59000
Lille, France
}%

\author{Henrik Bruus}
\email{bruus@fysik.dtu.dk}
\affiliation{%
 Department of Physics, Technical University of Denmark, DTU Physics Building 309,\\ 2800 Kongens Lyngby, Denmark
}%

\author{Michael Baudoin}
\email{michael.baudoin@univ-lille.fr}
\affiliation{%
Univ. Lille, CNRS, Centrale Lille, Univ. Polytechnique Hauts-de-France, UMR 8520 -
IEMN - Institut d’Electronique de Microélectronique et de Nanotechnologie, F-59000
Lille, France
}%

 \affiliation{Institut Universitaire de France, 1 rue Descartes, 75005 Paris}

\date{\today}

\begin{abstract}
Vortex-based single-beam tweezers have the ability to precisely and selectively move a wide range of objects, including particles, bubbles, droplets, and cells with sizes ranging from the millimeter to micrometer scale. In 2017, Karlsen and Bruus [Phys. Rev. Appl. 7, 034017 (2017)] theoretically suggested that these tweezers could also address one of the most challenging issues: the patterning and manipulation of miscible fluids. In this paper, we experimentally demonstrate this ability using acoustic vortex beams generated by interdigital transducer-based active holograms. The experimental results are supported by a numerical model based on acoustic body force simulations. This work paves the way for the precise shaping of chemical concentration fields, a crucial factor in numerous chemical and biological processes.
\end{abstract}

\maketitle


\section{\label{sec_intro}Introduction}


In both optics and acoustics, particle trapping and displacement using radiation forces were initially demonstrated with standing waves created by interfering counter-propagating waves. In 1986, Nobel Prize Laureate Arthur Ashkin demonstrated that 3D selective trapping of dielectric particles could be achieved with a \textit{single} focused laser beam \cite{ashkin1986observation}, ushering in a new era of optical manipulation. Inspired by this seminal work, Junru Wu \cite{wu1991acoustical} began exploring the trapping possibilities offered by focused beams in acoustics. He demonstrated the ability to trap particles using two opposing focused beams interfering near the focus to create a localized standing wave trap. Later, K.K. Shung and his team explored the potential of single focused beams for particle trapping \cite{lee2009single,zheng2012acoustic,chen2017adjustable,liu2017single}. However, as demonstrated later on by Gong and Baudoin \cite{gong2022single} a wide range of objects, including particles that are stiffer and denser than the surrounding fluids, can only be trapped laterally with a focused beam (no axial trap) and only in the Mie regime at specific frequencies near the particle resonances. Therefore, other methods are required to achieve 3D and robust trapping of such objects with a single beam.

Along these lines, Baresch, Marchiano, and Thomas first proposed theoretically \cite{baresch2013three} and later demonstrated experimentally \cite{baresch2016observation} that 3D trapping with single beams can be achieved using specific wavefields known as focused acoustical vortices, i.e. Bessel beams of topological order $l$ greater than one \cite{baudoin2020acoustic}. Marzo et al. \cite{marzo2015holographic} demonstrated in the Long Wavelength Regime (LWR), through an optimization algorithm, that these beams are optimal for single beam 3D trapping with the most even trapping capabilities along the three main axes. They also investigated the trapping capabilities of these beams (along with other types, such as twins or bottle beams) in air. Since then, the trapping capabilities of acoustic tweezers based on acoustical vortices have been extensively explored \cite{baudoin2020acoustic,guo2022review}. These tweezers have been shown to precisely and selectively trap and manipulate various objects, including bubbles \cite{baresch2020acoustic,lo2021tornado}, droplets \cite{lin2024selective}, kidney stones \cite{barnes2020sound}, microparticles \cite{riaud2017selective,baudoin2019folding,sahely2022ultra} and cells \cite{baudoin2020spatially}.

While the manipulation of objects with a clear interface has been widely demonstrated, the ability of single-beam tweezers to trap and manipulate miscible liquids with a diffuse interface, has not yet been demonstrated experimentally. Such manipulation presents a significant challenge due to (i) the generally weak contrast in density and compressibility between miscible liquids, (ii) the absence of an interface between the two fluids, which in the case of immisicible liquids contributes to the stiffness contrast of droplets through surface tension, and  (iii) the presence of diffusion, which severely limits the tweezers manipulation time. In other words, rapid manipulation is required before diffusion occurs, merges the two liquids, and reduces the contrast at the core of the acoustic force.

\begin{figure*}[t]
    \centering
    \includegraphics[width=\textwidth]{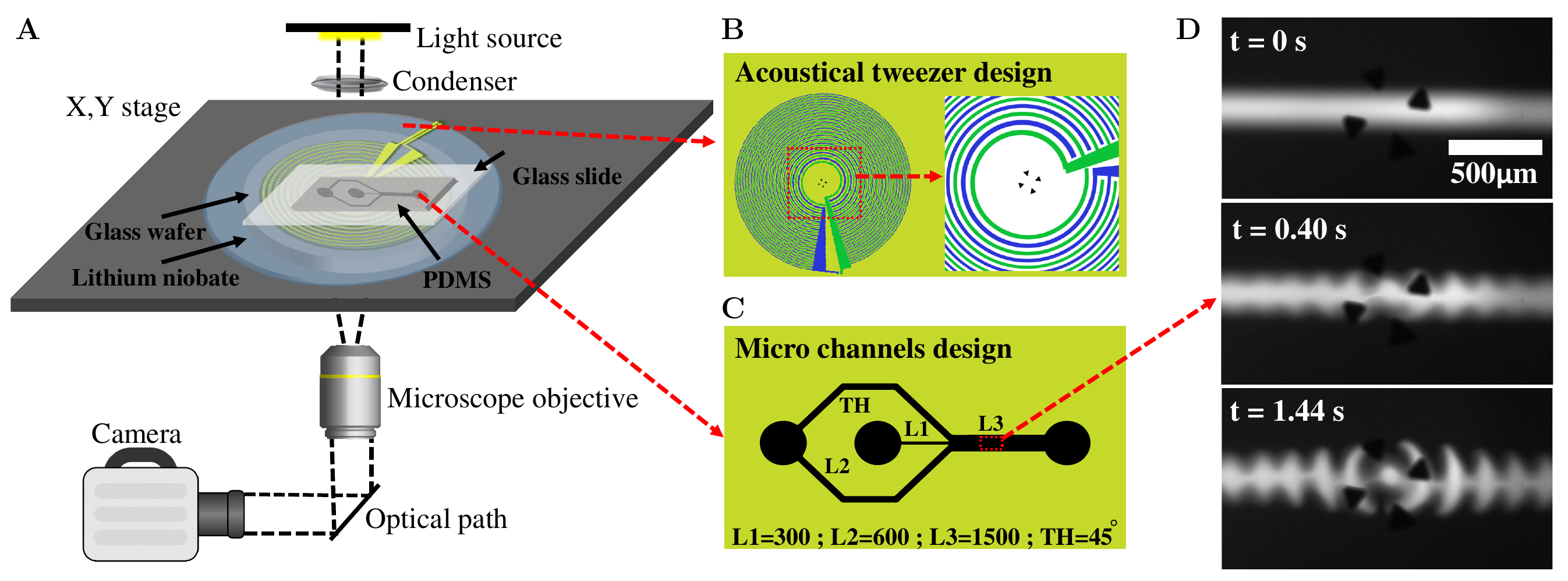}
    \caption{
    (A) Schematics of the experimentsl setup: the microchannel is placed on top of an acoustic tweezer, and in the lightpath of an optical microscope. The resulting image is recorded using a sCMOS camera.
    (B) Design of the tweezer, where the two electrodes are colored in blue and 
    green respectively.
    (C) Design of the co-flow system for creating parallel water and Ficoll lines; the channel of interest (L3)
    has a depth of \SI{40}{\micro\meter} and width of \SI{1500}{\micro\meter}.
    (D) Three example snapshots of the patterning process, which sets in upon
    turning on the acoustic tweezer.}
    \label{fig1_ExperimentalSetup}
\end{figure*}

The potential to manipulate inhomogeneous fluids with acoustic waves was explored experimentally by Deshmukh et al. \cite{deshmukh2014acoustic}. They demonstrated that acoustic forces generated by standing waves can modify miscible fluid distributions and stabilize density profiles against hydrostatic pressure gradients due to the higher impedance liquid being attracted to pressure nodes. They subsequently leveraged this discovery to introduce a new method, termed '\textit{isoacoustics}', for sorting cells based solely on their acoustic contrast, independently of their size \cite{augustsson2016iso,rezayati2023acoustophoretic,rezayati2023acoustofluidic}. The response of inhomogeneous fluids to acoustic fields was later rationalized by Karlsen et al. \cite{karlsen2016acoustic}, who introduced an acoustic force density model based on density and compressibility gradients to explain the forces acting on the fluid. Karlsen and Bruus further suggested, through numerical simulations \cite{karlsen2017acoustic}, the potential of acoustic tweezers based on focused beams or acoustical vortices to pattern and manipulate miscible fluids, although this was not demonstrated experimentally.

In this paper, we demonstrate the ability to pattern and manipulate miscible fluids using vortex-based acoustic tweezers synthesized with InterDigital Transducers (IDTs)-based active holograms. The experimental results are carefully compared with 2D numerical simulations based on the acoustic body force, showing good quantitative agreement.

\section{\label{sec_methodExp} Method: experiments}

\subsection{Setup description}

The experiments consist of two parts: (i) the creation of miscible fluid concentration patterns imposed by the vortex wavefield structure and (ii) the trapping and translation of a fluid blob trapped at the center of the vortex. These experiments were performed inside a \SI{40}{\micro\metre} deep and \SI{1.5}{mm} large microfluidic channel using an IDT-based tweezer (Fig.~\ref{fig1_ExperimentalSetup}.A). The tweezer (Fig.~\ref{fig1_ExperimentalSetup}.B) is connected to the bottom of the microfluidic chamber (made of a glass coverslip) using a drop of silicone oil. The result is visualized with a sCMOS high sensitivity Prime-BSI photometric camera operating at a steady frame rate of \SI{12.5}{fps} and imaging through a Nikon Ti2E optical microscope equipped with a module for fluorescent imaging.  A typical manipulation sequence consists of the following steps. (i) A centered narrow band of fluorescently labeled Ficoll solution ($\sim \SI{100}{\micro\metre}$ in width) surrounded by water is created using the co-flow system depicted on Fig.~\ref{fig1_ExperimentalSetup}.C, with two syringe pumps precisely controlling the fluids flow rates. (ii) The flow is stopped and the tweezer is activated. The tweezer creates a vortex beam, which is focused through a \SI{6.5}{mm} thick glass slide glued on top of the piezoelectric substrate, and then transmitted to the microfluidic chamber. This produces a pattern reminiscent of the vortex ring structure through respective motion of the two fluids (Fig.~\ref{fig1_ExperimentalSetup}.C). (iii) In the manipulation experiments, the blob of fluid trapped at the core of the vortex is displaced by moving the tweezer horizontally with a highly accurate XY Thorlabs motorized stage with a precision of \SI{100}{nm}. (iv) Eventually diffusion results in the merging of the two fluids and hence the disappearance of the concentration gradients.

\subsection{Acoustic tweezers design, fabrication and characterization}

\subsubsection*{\label{sec_design} Design}

The acoustic tweezers were designed using the method described in Refs.~\cite{baudoin2019folding,baudoin2020spatially}. In short, the phase map resulting from the intersection of a propagating (Hankel) acoustic vortex \cite{baudoin2020acoustic} of topological order 1 with the source plane is discretized over 4 levels $(0, \pi/2, \pi, 3\pi/2)$. The two levels corresponding to opposite phases $(0, \pi)$ are materialized into electrodes  sputtered on the surface of a \SI{0.5}{mm} thick, 3-inch Y-36 lithium niobate (LiNbO$_3$)  piezoelectric substrate, resulting in two intertwined spiralling lines of decreasing distance and width (see Fig.~\ref{fig1_ExperimentalSetup}.B). The two other phase levels are not materialized into electrodes, which enables to leave some space between the electrodes. 

In the present work, tweezers were designed to create a focused  vortex beam operating at the frequency $f=\SI{18}{MHz}$. A \SI{6.5}{mm} thick D263 glass wafer was glued on top of the electrodes, to ensure focusing (and hence better localization \cite{baudoin2019folding} of the trap) before the beam reaches the microfluidic chamber. The design of the electrodes was calculated to synthesize transverse waves inside the glass, as their speed of sound is closer to the one of water compared to longitudinal waves, hence ensuring  better wave transmission to the fluid contained in the microfluidic chamber. The transverse wave speed $c_t$ used to compute the propagation through the glass was chosen equal to \SI{3500}{\metre \second^{-1}}, which is an approximate value of the theoretical value calculated from the glass properties provided by the glass manufacturer using the formula $c_t = \sqrt{\frac{E}{2 \rho_g (1+\nu)}}$, with $E=\SI{72.9}{kN.mm^{-2}}$ the Young's modulus, $\nu = 0.208$ the Poisson's ratio and $\rho_g= \SI{2510}{kg.m^{-3}}$ the glass density. The spiralling structure was made of 16 turns, with a void central region with no electrodes to enable visualization through the tweezers. This resulted in a lateral radius of the patterned region of  $\sim$\SI{9}{mm} and hence an aperture of $\sim 54^\circ$.

\subsubsection*{Fabrication}

Classical photolithography technique was used to pattern spiraling metallic electrodes from metallic layers deposited on the substrate with Plasma Enhanced Vapor Deposition (PECVD): (i) The LiNbO$_3$ wafer undergoes ultrasonication in acetone and propanol for \SI{3}{min}, and is then dried with nitrogen gas. (ii) An adhesion promoter called HexaMethylDiSilazan (HMDS) is spread on the wafer using a spin coater, followed by a negative AZnLoF2020 photoresist with a thickness of \SI{3}{\micro\metre}. The resist is cured on a hot plate at $110^{\circ} \mathrm{C}$ for 90 seconds, after which the patterns of the tweezers are transferred into the resist using an optical mask and a MA6/BA6 SUSS Microtec UV optical aligner. (iv) The wafer is then heated on a hot plate at $110^{\circ} \mathrm{C}$ for \SI{120}{s} to complete the cross-linking process, and immersed in AZ326 developer for 30 seconds before being rinsed with deionized water.  (v) The wafer is coated with a layer of titanium (30 nm) and gold (30 nm) with PECVD. (vi) The lift-off process is achieved using a SVC14 solution at room temperature for one day, followed by ultrasonication at 35 kHz and $15 \%$ power to improve the efficiency of the lift-off operation.

A glass wafer of borosilicate D263 T with a diameter of 56.8 mm and a thickness of 6.5mm is then glued on top of the piezoelectric substrate using optically transparent epoxy glue (EPOTEK 301-2) to ensure good wave transmission. Prior to the glueing step, a \SI{15}{nm} thick chromium layer is deposited on the upper face of the glass wafer without etching to serve as markers for localizing the vortex center. The two substrates are cleaned with acetone and propanol and treated with $\mathrm{O}_2$ plasma to make the surfaces hydrophilic and improve glue spreading. The glue is degassed and mixed using an ARV 310 vacuum mixer to prevent bubble formation, and a drop of \SI{3.45}{\micro\liter}  glue is deposited on the center of the piezoelectric substrate. The glass wafer is then positioned atop the piezoelectric substrate and left on a horizontal plate until the glue covers the whole surface between the lithium niobate and glass substrates. After ensuring complete glue coverage, the structure is left to cure on the plate for two days at room temperature. The final step involves etching the markers on the glass wafer, which will enable to identify the vortex center in the experiments. The whole structure is cleaned with acetone and propanol and dried with nitrogen gas to remove dust, and the glass wafer is coated with AZ1505 resist (thickness about \SI{1}{\micro\metre}) using a spin coater and placed to dry for 1 day at room temperature. Once the patterns have been transferred onto the resist using the optical mask and MA6/BA6 SUSS aligner, the structure is immersed in MIF726 developer and subsequently washed with deionized water. The chromium layer is removed using a Cr etchant, and the tweezers are placed in acetone to remove resist traces, followed by propanol and drying with nitrogen gas.

\begin{figure}[b]
    \centering
    \includegraphics{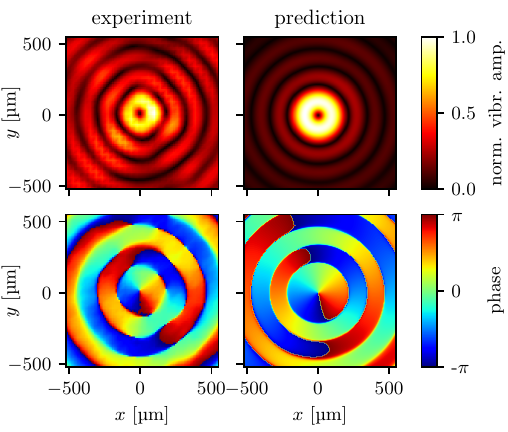}
    \caption{Vibrational field created by the tweezer
    driven at \SI{18}{MHz}. Left: Normalized amplitude (upper row) and phase (lower row) of the normal displacement measured experimentally with a UHF-120 Laser Doppler Vibrometer. Right: Normalized amplitude (upper row) and phase (lower row) of the field predicted by angular spectrum propagation of the electrodes binary source. Note that the maximum vibration amplitude is of the order of a few $\si{\nano\meter}$.}
    \label{fig2_AcousticField}
\end{figure}

\subsubsection*{Characterization}

The vibrational acoustic field produced by the tweezers was characterized using a Polytech UHF-120 Laser Doppler Vibrometer (LDV). The LDV measured the normal displacement  (amplitude and phase) at the bottom of the microfluidic chamber, i.e. at the top of the glass slide used as the support of the microfluidic PolyDiMethylSiloxane (PDMS) chamber. The wavefield measured experimentally represented on Fig.~\ref{fig2_AcousticField} is compared to simulations performed with Angular Spectrum (AS) code which takes the electrode design as a binary source plane. Both the amplitude and phase are faithful to the numerical predictions. The differences in the amplitude color map stems from slight under-sampling of the acoustic field which does not enable to capture precisely the maxima and minima (also clearly visible on the figure in Sec.~\ref{sec:num_conf}). The more pronounced experimental anisotropy might be related to the non-isotropic coupling coefficients along different directions owing to the crystal nature of the LiNbO$_3$ resulting in non isotropic amplitude along different directions as well as the presence of a set of two large electrodes (see Fig.~\ref{fig1_ExperimentalSetup}.B) intersecting the spiraling electrodes to efficiently deliver the current with minimal Joule losses.

Note that the two sets of electrodes of the tweezers are driven by a signal generator IFR 2023A providing a signal at \SI{18}{MHz} and \SI{5}{dBm} amplified with an AR50A250 \SI{150}{W} power amplifier, resulting in a substrate vibration with an amplitude of the order of \SI{1}{nm}. The precise value of the amplitude can vary according to the thickness of the coupling layer.

\subsection{Channel design and fabrication}

\subsubsection*{Design}

The microchannels are made of a PDMS slab fixed on top of a \SI{150}{\micro\metre} thick  glass slide. The same glass (Borosilicate D263) as the one used for the glass wafer was chosen as the bottom part of the channel to ensure good transmission of the wave between the glass wafer and the microchannel. PDMS was chosen to fabricate the microchannel due to its weak impedance mismatch with water and good absorbing properties, resulting in reduced resonance effects in the z direction. The channel design is represented on Fig.~\ref{fig1_ExperimentalSetup}.C and is similar to the one used in Ref.~\cite{deshmukh2014acoustic}. It is made of two inlets and a co-flow design which enables the creation of a central Ficoll line surrounded with water. The dimensions of the channel are given in Fig.~\ref{fig1_ExperimentalSetup}.C.  

\subsubsection*{Fabrication}

The PDMS slab is fabricated with the following procedure. (i) A mold is fabricated with lithography technique. To that end, a silicon wafer is first cleaned for \SI{15}{min} using piranha solution (mixture of sulfuric acid ($\mathrm{H}_2 \mathrm{SO}_4$) and hydrogen peroxide ($\mathrm{H}_2 \mathrm{O}_2$) in a 3:1 ratio). The wafer is then rinsed twice with de-ionized water and dried using nitrogen gas. After that, a negative photoresist, SU8-2035, is spin-coated onto the wafer to a thickness of \SI{40}{\micro\metre}, followed by baking at $65^{\circ} \mathrm{C}$ for \SI{3}{min}, and then at $95^{\circ} \mathrm{C}$ for \SI{6}{min}. To transfer the desired shape of the channel from an optical mask to the channel, the mask and wafer are aligned and exposed to UV light, after which the wafer is baked first at $65^{\circ} \mathrm{C}$ for \SI{2}{min}, and then at $95^{\circ} \mathrm{C}$ for \SI{6}{min}. The wafer is then treated with SU8 developer for \SI{5}{min}, rinsed with propanol, and dried using nitrogen gas. This ends up the fabrication of the mold required for PDMS patterning.

(ii) The mold is then placed in a 3-inch glass Petri dish and subsequently PDMS, mixed with a ratio 10:1 with the curing agent and blended to remove bubbles, is poured onto the mold. It is placed in a preheated oven at $110^{\circ} \mathrm{C}$ and left to cure for at least \SI{10}{min} until it solidifies. After cooling down, the PDMS is separated from the mold and the in- and outlets are punched using a biopsy puncher.

(iii) Finally, the PDMS slab is bonded to the glass cover slide.  To achieve this, the glass slide is cleaned with acetone and propanol, while the PDMS is rinsed with propanol and treated with oxygen plasma. Next, the PDMS is gently pressed onto the glass slide, followed by heating the assembly at \SI{70}{\celsius} for a minimum of \SI{10}{min} to ensure that the bond solidifies. The PDMS-glass assembly that results from this process is now ready to be used for microfluidic experiments.



\subsection{Ficoll preperation}

The Ficoll solution is made of a mixture of \SI{1.9}{g} of Ficoll (Cytiva Ficoll PM400), \SI{0.1}{g} fluorescent-labelled Ficoll (Sigma-Aldrich Polysucrose 400-fluorescein isothiocyanate conjugate) combined with \SI{11.33}{g} of deionized water to obtain a Ficoll solution with a mass concentration of 15\%. This concentration was chosen to obtain the best compromise between acoustic contrast, limited viscosity, and efficient fluorescence imaging. The chemicals were mixed using a magnetic stirrer bar inside a sealed Beaker for several hours until the Ficoll is fully dissolved.

\subsection{Post-processing}

To follow the miscible fluids dynamics induced by the activation of the tweezers, it is necessary to track the evolution of the Ficoll concentration. This concentration is tracked by monitoring light emission of fluorescent Ficoll. This fluorescent Ficoll has an excitation peak at \SI{496}{nm} and an emission peak at \SI{521}{nm}. Light emitted by the Nikon Intensilight C-HGFI Mercury Fluorescent light source and by the fluorescent Ficoll respectively were filtered by a GFP-4050B bandpass filter cube with excitation and emission bandwidth of $466 \pm 25$ nm and $525 \pm 25$ nm respectively and a dichroic mirror at \SI{495}{nm}. The filtered light emitted by the fluorescent molecule was then recorded using a high-sensitivity sCMOS Prime BSI camera.

\begin{figure}[t]
    \centering
    \includegraphics{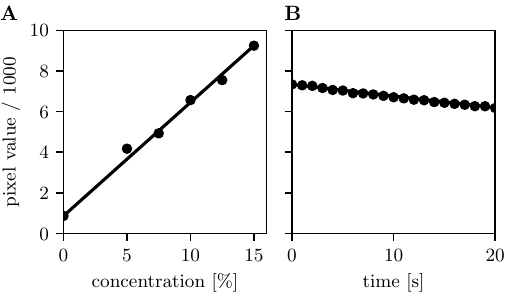}
    \caption{Typical trends (shown here for one single pixel) used for the calibration of the fluorescent imaging procedure. (A) Evolution of the grey scale pixel value captured by the camera as a function of the Ficoll concentration. (B) Variation of this grey value as function of time featuring a decrease of the fluorescent intensity with time.}
    \label{fig3_Postprocessing}
\end{figure}

The fluorescent emission captured by the camera was converted into concentration with the following calibration procedure: (i) Prior to the experiments, the micro-channel was filled successively with different solutions of Ficoll of controlled mass concentration ranging between 0\% and 15\%, obtained from the dilution of the initial solution described in the previous paragraph. (ii) Pictures of the channel filled with these different solutions were taken with the exact same settings (relative position of the channel and tweezers, light intensity, microscope and camera settings) as the one used for the final experiments. (iii) Due to a slight shadowing effect induced by the presence of the electrodes resulting in non uniform illumination, a calibration between the grey scale of the image taken by the camera and the concentration was performed for each pixel of the camera in the region of interest. A typical calibration curve is represented on Fig.~\ref{fig3_Postprocessing}.A, highlighting the linear relation between Ficoll concentration and the grey scale pixel value recorded with the camera. (iv) Finally, as we observed photobleaching (resulting in a decrease of fluorescence emission over time),  
the bleaching effect in presence of a quiescent liquid was evaluated for every pixel prior to experiments (see Fig.~\ref{fig3_Postprocessing}.B for an example), approximated by a linear function and subsequently corrected in the post-processing procedure.  To sum up, the post-processing procedure consists thus of a custom code translating grey-color values into fluorescent concentration taking into account the linear dependence of fluorescence on Ficoll concentration and correcting shading effects of the optical path as well as photobleaching over time. This calibration procedure was repeated for each set of experiments.

\begin{figure*}
    \centering
    \includegraphics{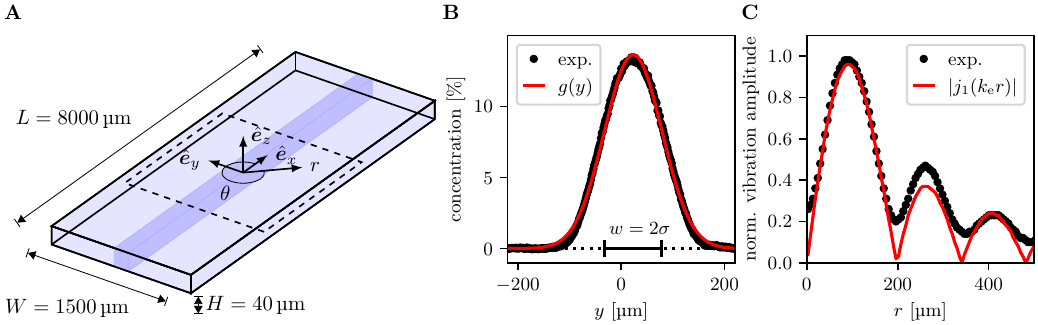}
    \caption{(A) Schematics (not to scale) of the simulated configuration. The dashed region indicates the computational domain of the 2D simulations.
    (B) Ficoll concentration along the channel width $y$. 
    A Gaussian $g(y)$ is fitted to the experimental data
    and used for the numerical simulations.
    (C) Normalized absolute vibration amplitude in the
    channel. The experimental data is averaged over the radial
    coordinate $r$. The Bessel function $ j_{1} (k_{\mathrm{e}} r) $ is fitted to the experimental values and used (with respective phase) for the numerical simulations, using a fitting parameter $ k_{\mathrm{e}} = \SI{24063}{\per\meter} $.}
    \label{fig3_NumericalSetup}
\end{figure*}

\section{\label{sec_methodNum}Method: Numerics}

A numerical model of the acoustic patterning is formulated based on the theory by Karlsen and Bruus \cite{karlsen2017acoustic}, in which acoustic fields acting on a short acoustic time scale $ t $ give rise to acoustic body forces $ \bm{f}_{\mathrm{ac}} $ on inhomogeneous fluid solutions on a slow time scale $ \tau $, due to density and compressibility gradients produced by spatially-varying solute concentration fields. By exploiting this disparity between $ t $ and $ \tau $ within the usual perturbative framework of theoretical acoustics, $ \bm{f}_{\mathrm{ac}} $ can be expressed as \cite{karlsen2016acoustic}

\begin{equation} \label{eq:fac}
    \begin{split}
        \bm{f}_{\mathrm{ac}} 
        & = - \frac{1}{4} \lvert p_{1} \rvert^{2} \bm{\nabla} \kappa_{0} - \frac{1}{4} \lvert v_{1} \rvert^{2} \bm{\nabla} \rho_{0} \\
        & = \frac{1}{4} \left(\kappa_{0} \lvert p_{1} \rvert^{2} - \rho_{0} \lvert \bm{v}_{1} \rvert^{2}\right) \frac{\bm{\nabla} \rho_{0}}{\rho_{0}} + \frac{1}{2} \kappa_{0} \lvert p_{1} \rvert^{2} \frac{\bm{\nabla} c_{0}}{c_{0}} \,,
    \end{split}
\end{equation}
where $ p_{1} $ and $ \bm{v}_{1} $ are the acoustic pressure and velocity fields, and $ \kappa_{0} $, $ \rho_{0} $, and $ c_{0} $ are the concentration-dependent compressibility, density, and speed of sound of an undisturbed fluid.

\subsection{\label{sec:num_conf}Description of the simulated configuration}

In accordance with the experiments, we consider a shallow microfluidic channel of length $ L = \SI{8000}{\micro\meter} $ along $ x $, width $ W = \SI{1500}{\micro\meter} $ along $ y $, and height $ H = \SI{40}{\micro\meter} $ along $ z $, featuring rigid walls and containing an inhomogeneous solution of Ficoll PM400 in water, as sketched in Fig.~\ref{fig3_NumericalSetup}.A. The local concentration of Ficoll is represented by a mass fraction field $ s (\bm{r}, \tau) $, which initially is a straight band 
along the $ x $-axis, constant along $ z $, and with a Gaussian smearing across the width in $ y $,
\begin{equation} \label{eq:sinit}
    s (\bm{r}, \tau = 0) = s_{\mathrm{max}} \exp{\left[-\frac{1}{2} \left(\frac{y}{\sigma_{0}}\right)^{2}\right]} \, ,
\end{equation}
where $ s_{\mathrm{max}} $ and $ \sigma_{0} $ are determined from best fit to experimental data (Fig.~\ref{fig3_NumericalSetup}.B), which, following the procedure presented in the supplementary material of Ref.~\cite{augustsson2016iso}, leads to the constant diffusivity$ D_{0}$. Furthermore, the solution is characterized by an $ s $-dependent mass density $ \rho_{0} (s) = \rho_{\mathrm{w}} (1 + a_{1} s) $, speed of sound $ c_{0} (s) = c_{\mathrm{w}} (1 + c_{1} s) $, and viscosity $ \eta_{0} (s) = \eta_{\mathrm{w}} \exp{(b s)}$, as in Ref.~\cite{Qiu2019}.

An acoustic vortex travels up though the glass and is then transmitted to the fluid of the channel. Owing to (i) the weak acoustic contrast between water and PDMS and strong absorbing properties of the PDMS  resulting in weak reflection, (ii) the small depth of the channel compared to the attenuation length ($\sim \SI{14}{cm}$ in water at \SI{18}{MHz}), the field intensity variations along the $z$ axis are neglected. Hence, the acoustic wave pressure  lateral profile $ p_{1}(r) $ in the fluid is assumed to be the same profile as the one measured with LDV  at the bottom of the channel well approximated by the function (see Fig.~\ref{fig3_NumericalSetup}.C),
\begin{equation} \label{eq:ac}
    p_{1} = p_{\mathrm{a}} \frac{j_{1} (k_\mathrm{e} r)}{j_{1}^{\mathrm{max}}} \mathrm{e}^{\mathrm{i} (\theta - \omega t)},
\end{equation}
where $ (r,\theta, z) $ are the cylindrical polar coordinates with its origin at the center point of the channel, $ p_{\mathrm{a}} $ is the acoustic pressure amplitude, $ k_e $ an effective wavenumber, $ \omega = 2 \pi f $ the angular frequency, $ j_{1} $ the spherical Bessel function of order one, and $ j_{1}^{\mathrm{max}} = 0.436 $ the maximum of $ j_{1} $. The consistency of such approximation has been verified in ref. \cite{baudoin2020spatially} through angular spectrum computation of the transmission of the acoustic field from the glass slide to the fluid.
The corresponding acoustic velocity $ \bm{v}_{1} $ is then calculated using the first-order momentum equation,
\begin{equation} \label{eq:acbis}
 \quad \bm{v}_{1} = - \frac{\mathrm{i}}{\omega \rho_{0}} \bm{\nabla} p_{1} \,.
\end{equation}
 Note that if a spherical vortex was generated with transducers positioned all around the focal point, the effective wavenumber $ k_{\mathrm{e}} $ would correspond to the actual wavenumber $ k = \omega / c_{\mathrm{t}} $. But the finite aperture (here $54^\circ$) limits lateral focusing resulting in larger rings. Hence, this parameter was optimized to obtain the best match with experiments (Fig.~\ref{fig3_NumericalSetup}.C), leading to the value $ k_{\mathrm{e}} = \SI{24063}{\per\meter} $. 

The equations governing the slow time-scale hydrodynamic velocity $ \bm{v} (\bm{r}, \tau) $, pressure $ p (\bm{r}, \tau) $, and Ficoll concentration $ s (\bm{r}, \tau) $ consist of the Navier--Stokes equation with $ \bm{f}_{\mathrm{ac}} $ as a body force, the continuity equation, and the advection-diffusion equation,
\begin{subequations} \label{eq:3Deqs}
    \begin{align}
        \rho_{0} \left[\partial_{\tau} \bm{v} + (\bm{v} \cdot \bm{\nabla}) \bm{v}\right] & = \bm{\nabla} \cdot \bm{\sigma} + \bm{f}_{\mathrm{ac}} \,, \\
        \partial_{\tau} \rho_{0} + (\bm{v} \cdot \bm{\nabla}) \rho_{0} & = - \rho_{0} \bm{\nabla} \cdot \bm{v} \,, \\
        \partial_{\tau} s + (\bm{v} \cdot \bm{\nabla}) s & = \bm{\nabla} \cdot (D_{0} \bm{\nabla} s) \,.
    \end{align}
\text{Here, $ \bm{\sigma} $ is the fluid stress tensor, given by}
    \begin{equation}
    \bm{\sigma} = - p \bm{I} + \eta_{0} \left[ \bm{\nabla} \bm{v} + (\bm{\nabla} \bm{v})^{\mathsf{T}} \right] + (\eta_{0}^{\mathrm{b}} - \tfrac{2}{3} \eta_{0}) (\bm{\nabla} \cdot \bm{v}) \bm{I} \,,
\end{equation}
\end{subequations}
where $ \bm{I} $ is the identity tensor, $ (\cdot)^{\mathsf{T}} $ indicates the transpose, and $ \eta_{0} $ and $ \eta_{0}^{\mathrm{b}} \approx 2.79 \eta_{0} $ (valid for pure water) are the shear and bulk viscosities, respectively. These governing equations are supplemented with no-slip and no-flux conditions on the rigid walls,
\begin{equation} \label{eq:3DBCs}
    \bm{v} = \bm{0} \,, \quad (\hat{\bm{n}} \cdot \bm{\nabla}) s = 0 \,,
\end{equation}
with $ \hat{\bm{n}} $ being the outwards pointing unit normal to the boundary surface. 

We have numerically simulated this 3D model; however, memory requirements prevent us from simulating it at the actual size of the system. We hence proceeded by creating a reduced 2D model, obtained by averaging the 3D model over the height dimension along $ z $, which approximates the 3D model well in the limit of long acoustic wavelengths compared to the channel height, as quantified by the dimensionless parameter $ \epsilon = (k_{\mathrm{e}} H / 2)^{2} \ll 1 $. In experiments, we find that $ \epsilon = 0.23 $.

\begin{figure*}[t]
    \centering
    \includegraphics{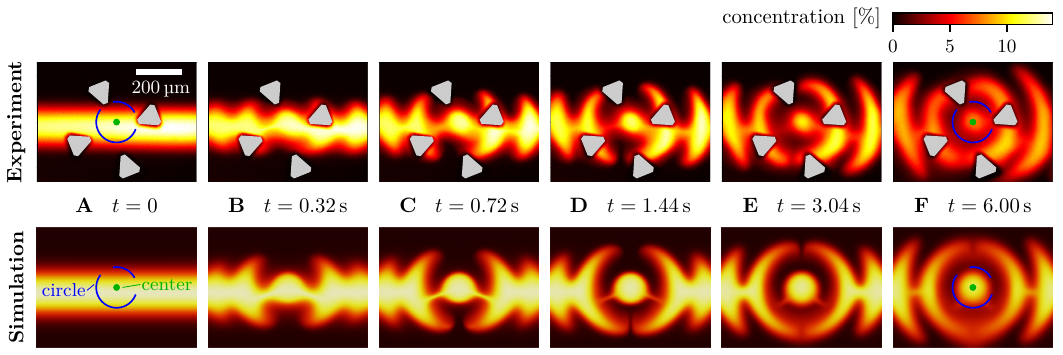}
    \caption{Snapshots of the experimental (top row) and simulated (bottom row) evolution of Ficoll concentration after activation of the tweezer. The reader may refer to the videos available in the supplementary material (Movie S1). The grey triangles in the upper, experimental row correspond to the markers of the tweezer, which are aligned around it's center and black in the original snapshots. The circle (blue) and point (green) used to define the normalized concentration difference $\Delta \tilde{s}$ of Eq.~\ref{eq:DeltacDef} between the first ring and the center of the vortex are shown. The colour bar defines the concentration range in the snapshots.}
    \label{fig_evolutionInTime}
\end{figure*}

The 2D height-averaged model is formulated in terms of the height-averaged fields $ \langle \bm{v} \rangle $, $ \langle p \rangle $, and $ \langle s \rangle $, defined by
\begin{equation} \label{eq:heightaverage}
    \langle f \rangle (x,y,\tau) \equiv \frac{1}{H} \int_{-H/2}^{H/2} f (x,y,z,\tau) \, \mathrm{d} z \,.
\end{equation}
When $ \epsilon \ll 1 $, the acoustic-induced spatial variations in the lateral dimensions, such as the concentration field pattern and underlying flow fields, evolve significantly slower than those in the height dimension. As a result, the velocity field adopts a parabolic profile in the height dimension, $ \bm{v} = \tfrac{3}{2} \langle \bm{v} \rangle \left[ 1 - (2 z / H)^{2} \right] $, and the concentration remains largely uniform along $ z $, with only a small variation caused by the parabolic velocity profile, which leads to classical Taylor dispersion. This enables us to height-average the governing equations \eqref{eq:3Deqs}, resulting in the following 2D equations,
\begin{subequations} \label{eq:GoverningEqs}
    \begin{align}
        \eta_{0} \nabla_{\parallel}^{2} \langle \bm{v} \rangle - \frac{12 \eta_{0}}{H^2} \langle \bm{v} \rangle & = \bm{\nabla}_{\parallel} \langle p \rangle - \bm{f}_{\mathrm{ac}} \left[ \bm{\nabla}_{\parallel} \langle s \rangle \right] \,, \\
        0 & = \bm{\nabla}_{\parallel} \cdot \langle \bm{v} \rangle \,, \\
        \partial_{\tau} \langle s \rangle + (\langle \bm{v} \rangle \cdot \bm{\nabla}_{\parallel}) \langle s \rangle & = \bm{\nabla}_{\parallel} \cdot (\bm{D}_{0} \cdot \bm{\nabla}_{\parallel} \langle s \rangle) \,,
    \end{align}
\end{subequations}
where $ \bm{\nabla}_{\parallel} = \hat{\bm{e}}_{x} \partial_{x} + \hat{\bm{e}}_{y} \partial_{y} $ is the lateral part of the gradient $ \bm{\nabla} $, and $ \bm{D}_{0} = D_{0} \bm{I} + \tfrac{1}{210} \frac{H^2}{D_{0}} \langle \bm{v} \rangle \langle \bm{v} \rangle $ is the dispersion tensor, valid in the long-wavelength limit (where $ \langle \bm{v} \rangle $ varies slowly) and for sufficiently high Péclet numbers, $ Pe = \tfrac{\lvert \langle v \rangle \rvert H}{2 D_{0}} \gg 1 $. The term $ - \tfrac{12 \eta_{0}}{H^{2}} \langle \bm{v} \rangle $ represents the viscous damping exerted by the top and bottom walls. The height-averaged versions of Eq. \eqref{eq:3DBCs} are
\begin{equation}
    \langle \bm{v} \rangle = \bm{0} \,, \quad (\hat{\bm{n}} \cdot \bm{\nabla}_{\parallel}) \langle s \rangle = 0 \,.
\end{equation}

This 2D height-averaged model has been simulated and successfully compared with the full 3D model for a down-sized version of the system, where the 3D model is tractable, see Appendix~\ref{Sec:comp2D3D}.

\subsection{Description of the numerical method}

\begin{table}[b]
    \centering
    \caption{Parameter values used in the numerical simulations. The frequency $ f $ is given in Sec.~\ref{sec_design}, the effective wavenumber $k_{\mathrm{e}}$ and the diffusivity $ D_{0} $
    are obtained from fitting to experiments (see Fig.~\ref{fig3_NumericalSetup}), and the remaining parameter values are from Table II in Ref.~\cite{Qiu2019}.}
    \begin{tabular}{lrllrl}
    \toprule
    \multicolumn{3}{c}{\textit{Acoustics}} & \multicolumn{3}{c}{\textit{Ficoll solution}} \\
    \cmidrule(lr){1-3}
    \cmidrule(lr){4-6}
    \multicolumn{1}{l}{Parameter} & \multicolumn{1}{c}{Value} &
    \multicolumn{1}{c}{Units} &
    \multicolumn{1}{l}{Parameter} & \multicolumn{1}{c}{Value} &
    \multicolumn{1}{c}{Units} \\
    \midrule
    $ f $ & $ 18.0 $ & \si{\mega\hertz} & $ D_{0} $ & $ 9.19 \times 10^{-11} $ & \si{\square\meter\per\second} \\
    $ k_{\mathrm{e}} $ & $ 24063 $ & \si{\per\meter} & $ \rho_{\mathrm{w}} $ & 996.85 & \si{\kilo\gram\per\cubic\meter} \\
    $ \epsilon = \left(\tfrac{k_{\mathrm{e}} H}{2}\right)^{2} $ & $ 0.23 $ &  & $ c_{\mathrm{w}} $ & 1496.30 & \si{\meter\per\second} \\
      &  &  & $ \eta_{\mathrm{w}} $ & 0.893 & \si{\pascal\second} \\
    & & & $ a_{1} $ & 0.349 & \\
    & & & $ c_{1} $ & 0.167 & \\
    & & & $ b $ & 16.20 & \\
    \bottomrule
    \end{tabular}
    \label{tab:paramval}
\end{table}

Numerical simulations are carried out in the finite element software COMSOL Multiphysics \cite{comsol}, following a similar procedure to that of Ref.~\cite{karlsen2017acoustic}, including a standard mesh-convergence analysis \cite{Muller2012} (Appendix~\ref{sec_meshConv}). The acoustic fields in Eq. \eqref{eq:ac} are implemented analytically through $ \bm{f}_{\mathrm{ac}} $, and the concentration $ s $ is represented by a logarithmic field, $ \ln{(s)} $, to improve numerical stability. The computational domain $ \Omega $ has been reduced to the square subdomain indicated by the dashed lines in Fig.~\ref{fig3_NumericalSetup}.A by exploiting the rapid radial decay of $ \bm{f}_{\mathrm{ac}} $, which ensures that the associated error of introducing the artificial boundaries is sufficiently small. Symmetries have been exploited to reduce the computational domain further when appropriate. The domain is meshed with free triangular elements, whose sizes increase with distance from the acoustic center. The parameters used for simulations are summarized in Table \ref{tab:paramval}.

The time-dependent equations are solved using a backward-differentiation time-stepping method with outputs every \SI{0.04}{\second} and a strict adaptive stepping scheme. The cell Peclet number $ Pe_{\mathrm{cell}} $ and the total amount of solute, $ S = \int_{\Omega} s \, dV $, are monitored at all times during simulations to ensure that $ Pe_{\mathrm{cell}} $ remains reasonably small and that $ S $ is conserved.

\section{\label{sec_resultsSingle}Results}

\subsection{\label{sec_miscFluids}Miscible fluids patterning}

A typical sequence of the dynamic evolution of the Ficoll concentration field upon activation of the tweezer is presented in Fig.~\ref{fig_evolutionInTime} (top row) and compared to numerical simulations (bottom row). As expected from the density and compressibility contrasts between Ficoll and water, Ficoll is moved toward the minima of the pressure field and water to the maxima (A to D) by the acoustic body force, ending up in the formation of ring structures reminiscent of the vortex amplitude map (Fig. \ref{fig2_AcousticField}). However, as diffusion becomes dominant, it leads to progressive homogenization of the two fluids and blurred rings (E to F). Indeed, as the acoustic body force is proportional to the gradient of the fluid acoustic properties (density, compressibility) it cannot prevent diffusion. The main features of this time sequence are well captured by the numerical simulations  (Fig.~\ref{fig_evolutionInTime}, bottom row). 

\begin{figure} [t]
    \centering
    \includegraphics{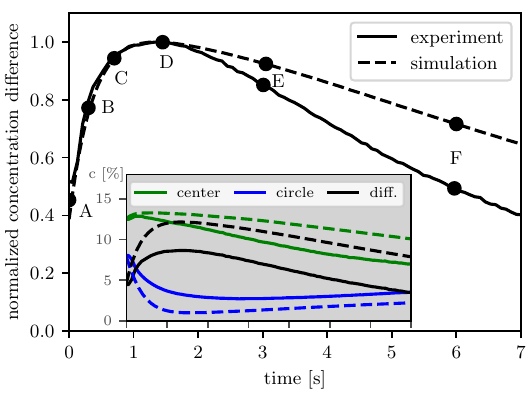}
    \caption{Plot of the time evolution of the normalized concentration difference $\Delta \tilde{s}$ of Eq.~(\ref{eq:DeltacDef}). Solid lines and dashed lines indicate experimental and numerical results respectively. The labeled dots correspond to the snapshots represented in \cref{fig_evolutionInTime}. The inset shows the time evolution of the corresponding center concentration $s_\mathrm{cent}$  (green), the concentration $\langle s \rangle_\mathrm{circ}$ averaged over the first maximum intensity ring (blue), and the difference between the two (black). The locations of the center and ring circle chosen for the evaluation of the concentration are represented in green and blue respectively on Fig.~\ref{fig_evolutionInTime}. }
    \label{fig_TimeVsAmplitude}
\end{figure}

For a more quantitative comparison of the evolution of the concentration gradients between regions of maximum and minimum acoustic intensity, we plotted on Fig.~\ref{fig_TimeVsAmplitude} (insert) the evolution of the Ficoll concentration $\langle s \rangle_\mathrm{circ}$ averaged over the maximum intensity ring (represented by a blue circle  on Fig.~\ref{fig_evolutionInTime}) and at the vortex center  $\langle s \rangle_\mathrm{cent}$ (green point on Fig.~\ref{fig_evolutionInTime}). Since the absolute values of the concentration are very sensitive to the exact position of the circle and center, and the relative position of the tweezers compared to the central Ficoll line, we further compared the normalized concentration difference $\Delta \tilde{s}$ between the circle and center concentration,
\begin{equation}
\label{eq:DeltacDef}
\Delta \tilde{s} = \frac{\langle s \rangle_\mathrm{circ}-\langle s \rangle_\mathrm{cent}}{\mathrm{max}(\langle s \rangle_\mathrm{circ}-\langle s \rangle_\mathrm{cent})}.
\end{equation}
We have plotted $\Delta\tilde{s}$, $\langle s \rangle_\mathrm{cent}$, and $\langle s \rangle_\mathrm{circ}$ in Fig.~\ref{fig_TimeVsAmplitude}, where the labels A to F corresponds to the image sequence of Fig.~\ref{fig_evolutionInTime}. The curves exhibit a rapid increase in the concentration gradient upon activation of the tweezers (A to D) and then a slow decrease due to the diffusion process (D to F). While the first part of the sequence is very well captured by the predictions of the numerical simulations, some deviations appear once the concentration difference has reached a maximum. A comparison between the two curves indicates that the diffusion process is undervalued by the simulations in this second part of the sequence. These deviations might stem from the exclusion in the numerical models of (i) thermal processes, which could affect the fluid dynamics as well as photobleaching, and (ii) acoustic streaming, which could create some additional recirculating flows accelerating the fluid merging process.

\begin{figure}[t]
    \centering
    \includegraphics{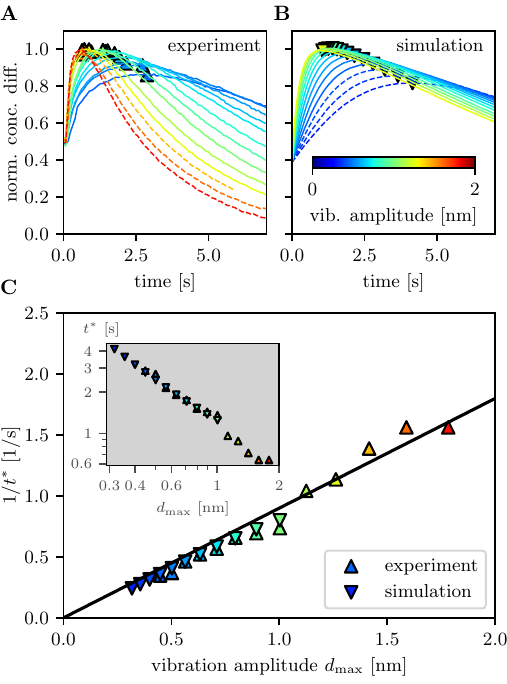}
    \caption{(A) Experimental and (B) numerical normalized concentration difference $\Delta \tilde{s}$ as a function of time. The curves are normalized by the maximum overall concentration for the experiments and numerical simulations respectively. Each line corresponds to a given vibration amplitude (see scalebar). Dashed lines indicate amplitudes that are only available numerically (low amplitudes) or experimentally (high amplitudes). The markers correspond to the maxima of these curves, which we define as the characteristic time $t^\ast$.
    (C) Inverse of the characteristic times $t^\ast$ as a function of the vibration amplitude.
    The inlet shows the same data but with $t^\ast$ and in a logarithmic scale.}
    \label{fig6_TimeVsAmplitude}
\end{figure}

\subsection{Characteristic time of pattern formation}

As it is a crucial parameters for applications, we further investigated the evolution of the time $t^*$ required to reach the maximum concentration difference between the vortex center and the first minimum intensity as a function of the wave amplitude (Fig.~\ref{fig6_TimeVsAmplitude}). Both the experiments and the simulations exhibit a relatively linear decrease (exponent $-1.09$ in the log-log plot for best fit with the combined experimental and numerical data) of the time  $t^*$ with the applied vibration amplitude. 

Note that the experimental wave normal amplitude at the beam center was characterized with the LDV and showed a linear dependence with the electrical power applied by our signal generator. Hence, the relative evolution of the vibration amplitude in a set of experiments is known. Nevertheless, the absolute value of the vibration amplitude shall depend on the thickness of the coupling silicone oil layer, which might differ between the LDV measurements and a set of experiments. Hence the absolute vibration amplitude was obtained in Fig.~\ref{fig6_TimeVsAmplitude} by determining the best match between experimental  and numerical values of $t^*$, and assuming a relation $p_1 \approx \rho_w c_w \omega d_1^n$ between the normal vibration amplitude $d_1^n$ and the pressure field $p_1$. The range of vibration amplitudes obtained in this way is of the same order of magnitude as the vibration amplitude measured with the LDV.

\begin{figure}[t]
    \centering
    \includegraphics{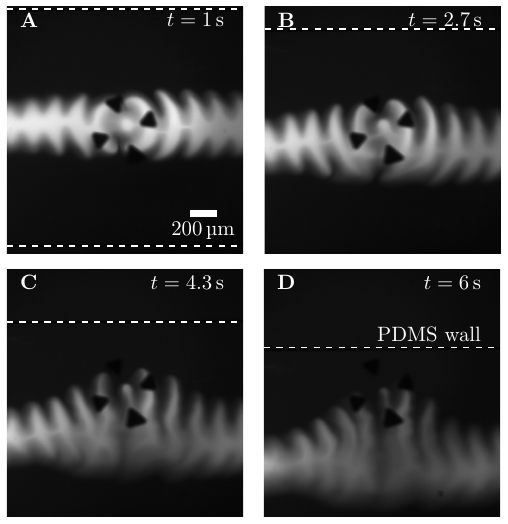}
        \caption{Translation of a blob of Ficoll solution (here with a concentration of $20 \%$ in mass) trapped at the center of an acoustic vortex. The channel is displaced with respect to the tweezer \SI{1.1}{s} after turning on the tweezer. The channel walls of the PDMS channel are indicated as reference of the displacement undergone.}
    \label{fig8_displacement}
\end{figure}

\subsection{Experimental manipulation of a Ficoll blob}

To conclude this work, we further explored the possibility to move the blob of fluid trapped at the center of  acoustic vortex after formation of the patterns. Fig.~\ref{fig8_displacement} and movie S2 show translation along the $y$ axis and movie S3 shows the successive displacement along the two orthogonal directions x and y. These results constitute the first experimental evidence of a trapped fluid blob translation with vortex-based tweezers as theoretically predicted by Karlsen and Bruus \cite{karlsen2017acoustic}. The translational capability could be further improved in the future, by employing fluid combinations with stronger acoustic contrast and  tweezers with weaker secondary rings. Note that in this section, the results have been obtained with a $20\%$ mass concentration Ficoll solution. 

\section{Conclusion, perspectives}

In this paper we demonstrated the ability to pattern and manipulate inhomogeneous miscible fluids using acoustic tweezers based on interdigital transducers. The experimental trends are recovered through numerical simulations based on the formulation proposed by Karlsen et al. \cite{karlsen2016acoustic}, which relies on the acoustic force density. Besides the precise manipulation of fluid blobs with selective acoustical tweezers, the development of a wealth of holographic techniques -- passive \cite{melde2016holograms,jiang2016convert,jimenez2016formation,jimenez2018sharp,jimenez2019holograms,ma2020acoustic,li2021three,xu2023sound,xu2024micro}, active \cite{riaud2017selective,muelas2018generation,baudoin2019folding,baudoin2020spatially,muelas2020active,sahely2022ultra} and dynamic \cite{riaud2015anisotropic,riaud2015taming,baresch2016observation,riaud2016saw,marzo2019holographic,ma2020spatial}) -- paves the way toward complex patterning of chemical and biomolecule gradient concentrations within microfluidic chambers. This opens a wide range of perspectives ranging from controlled studies of the fundamental roles played by biomolecular gradients in various biological processes \cite{keenan2008biomolecular} to the design of reconfigurable microlens arrays \cite{gao2022demand}.
{\color{blue}
\section*{}
}

\appendix

\section{\label{Sec:comp2D3D}Numerical validation of the height-averaged 2D model}

To validate the height-averaged 2D model, we have compared it to a full 3D model. However, due to computer memory requirements, this numerical comparison has been carried out on a reduced size of the system, namely a cylindrical subdomain with the correct height $ H $, but with a small radius $ R = \SI{444}{\micro\meter} $. The radius approximately coincides with a node in the fluid velocity field to minimize the error from the artificial rigid boundary. The axis of the cylinder goes through the center of the acoustic wave, and the initial Ficoll band coincides with the center. All of these measures were necessary to enable the 3D simulations, which ultimately took several days to complete.

\begin{figure}[t]
    \centering
    \includegraphics{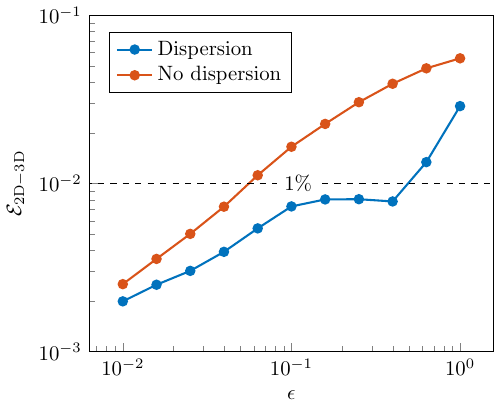}
    \caption{Convergence of the 2D model versus $ \epsilon = (k_\mathrm{e} H / 2)^{2} $. The error $\mathcal{E}_\mathrm{2D-3D}$ is calculated as the normalized $L^{2}$-norm difference between the concentration difference curves obtained by the 2D model (with and without Taylor dispersion) and the 3D model. As $ \epsilon $ decreases, better agreement between 2D and 3D models is attained.}
    \label{fig:EpsilonConvergence}
\end{figure}

Figure~\ref{fig:EpsilonConvergence} shows the normalized $ L^{2} $-norm difference $ \mathcal{E}_\mathrm{2D-3D} $ between the concentration difference curves $ \Delta \tilde{s}_{\mathrm{2D}} $ and $ \Delta \tilde{s}_{\mathrm{3D}} $ obtained by the 2D height-averaged model (both with and without Taylor dispersion) and the 3D model over the slow time variable $ \tau $, as a function of $ \epsilon = (k_\mathrm{e} H / 2)^{2} $,
\begin{equation}
    \mathcal{E}_\mathrm{2D-3D} = \sqrt{\frac{\int \lvert \Delta s_\mathrm{2D} - \Delta s_\mathrm{3D} \rvert^{2} \, \mathrm{d} \tau}{\int \lvert \Delta s_\mathrm{3D} \rvert^{2} \, \mathrm{d} \tau}} \,.
\end{equation}
The error of the height-averaged 2D model decreases with decreasing $ \epsilon $. Furthermore, the 2D model with Taylor dispersion consistently produces smaller errors than the 2D model without Taylor dispersion. The parameter $ \epsilon $ was varied by changing $ H $, and to compensate for the increased viscous damping at small $ \epsilon $ and maintain the Péclet number, the acoustic body was increased by the square of the relative change in $ H $ from the default value.

\section{\label{sec_meshConv} Mesh convergence study}
A standard mesh-convergence study was conducted to ensure that simulations results are independent of the mesh \cite{Muller2012}. Solutions were computed on increasingly refined meshes, each identified by a mesh refinement factor $ CONV $, which reduces the maximum mesh element size for increasing values. The normalized $L^{2}$-norm difference $ \mathcal{E} $ was calculated for the solution $ u $ for a given value of $ CONV $, using the solution on the most refined mesh as the reference $ u_{\mathrm{ref}} $,
\begin{equation}
    \mathcal{E}_\mathrm{mesh} = \sqrt{\frac{\int_\Omega \lvert u - u_{\mathrm{ref}} \rvert^{2} \, \mathrm{d} V}{\int_\Omega \lvert u_{\mathrm{ref}} \rvert^{2} \, \mathrm{d} V}} \,.
\end{equation}

\begin{figure}[h]
    \centering
    \includegraphics{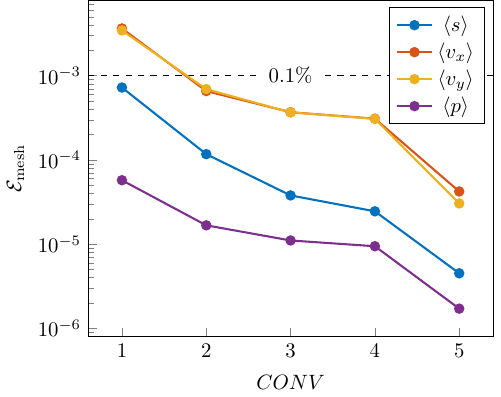}
    \caption{Time-averaged normalized $ L^{2} $-norm difference between solutions of the 2D height-averaged model on increasingly refined meshes, which are identified by increasing values of $ CONV $, and the solution on the most refined mesh.}
    \label{fig:MeshConvergence}
\end{figure}

Figure \ref{fig:MeshConvergence} shows $ \mathcal{E} $ vs.\ $ CONV $, time-averaged over 0 to \SI{5}{\second} for the 2D height-averaged model for $ d_{1}^{n} = \SI{0.5}{\nano\meter} $. For the final simulations, we used the mesh corresponding to $ CONV = 2 $, at which numerical errors become smaller than $ 0.1 \% $.

\section*{Acknowledgement}

The authors acknowledge the support of the International Chair Will ICMINA, financed by Agence Nationale de la Recherche "ANR-21-IDES-0006". 


%

%

\end{document}